\documentclass[manuscript]{aastex}
\input epsf
\usepackage{graphics}
\begin{document}
\title{Ground-based near-infrared imaging of the HD141569 circumstellar disk}
\author{A. Boccaletti}
\affil{CEA/Saclay, Service d'Astrophysique, France \\
\& California Institute of Technology, Pasadena CA, USA}
\author{J.-C. Augereau}
\affil{CEA/Saclay, Service d'Astrophysique, France}
\author{F. Marchis}
\affil{University of California, Berkeley CA, USA}
\author{J. Hahn}
\affil{Lunar and Planetary Institute, Houston TX, USA }

\shorttitle{First ground-based imaging of the HD141569 circusmtellar disk}
\shortauthors{Boccaletti et al.}
\received{} \revised{} \accepted{}
\begin{abstract}
We present the first ground-based near-infrared image of the circumstellar disk
around the post-Herbig Ae/Be star HD141569A initially detected with
the HST. Observations were carried out in the near-IR (2.2\,$\mu$m) at
the Palomar 200-inch telescope using the adaptive optics system PALAO.
The main large scale asymmetric features of the disk are detected on
our ground-based data. 
In addition, we measured that the surface brightness of the
disk is slightly different than that derived by HST observations (at
1.1\,$\mu$m and 1.6\,$\mu$m). 
We interpret this possible color-effect in terms of dust properties
and derive a minimal grain size of $0.6\pm0.2\mu$m for compact grains and a power law index for the grain size distribution smaller than $-3$.
Basic dynamical considerations are consistent with the presence of a remnant amount of gas in the disk.
\end{abstract}
\keywords{stars: circumstellar matter, stars: individual HD141569A,
techniques: high angular resolution, techniques: image processing}
\newpage
\section{Introduction}
A circumstellar disk around the star HD141569A was detected
independently by Augereau et al. (1999) and Weinberger et al. (1999)
using the near-IR camera NICMOS2 on the HST. These first images
revealed an optically thin annular disk with a complex morphology: 1/
an annular structure peaking at 325AU evidenced at 1.1\,$\mu$m and
1.6$\,\mu$m; 2/ a second axisymmetrical ring at 185AU unambiguously
detected at 1.1\,$\mu$m and less obvious in 1.6$\,\mu$m images.  More
recent data obtained with STIS in the visible confirm the presence of
two grain populations but also exhibit some strong asymmetries,
especially in the closer annular pattern (arc-like patterns for
instance), with an unprecedented angular resolution (Mouillet et
al. 2001).  According to Mouillet et al. (2001), the anisotropic
scattering of the light cannot account itself for asymmetric features
and one has to invoke a non axisymmetrical distribution of dust very
likely provided by the gravitationnal influence of pertubator(s)
within the disk (an hypothetic planetary companion for instance),
outside the disk (the two stellar companions lying at 7.54'' and
8.93'' from the star, 1065\,AU and 1370\,AU respectively if the
companions are in the disk plane), or both. Some numerical simulations
are ongoing to take into account the influence of the two T-Tauri like
companions (Augereau et al., in preparation). \\ In addition, a close
mid-IR emission previously inferred by spectral energy distribution
fitting (Augereau et al. 1999) was detected within 100\,AU (Fisher et
al. 2000) with the Keck telescope at $10.8\,\mu$m and $18.2\,\mu$m.
The complex environment of HD141569A clearly indicates that this star
is in an evolved stage of evolution and may have already experienced a
planetary formation stage.

We used the star HD141569A as a fiducial case to assess the capability
of the adaptive optics system at the Palomar 200-inch telescope for
detecting circumstellar disks around relatively young stars.
Consequently, we report the first ground-based imaging of this disk in
scattered light at near IR wavelengths. Despite a much lower
sensitivity than the HST, our data feature a better angular resolution
for identical wavelengths.

The observing sequence and the data reduction process are detailed in
section 2 and the subsequent results are analysed in section 3. We
discuss grains properties in section 4.
\section{Observation and Data Reduction}
Observations of HD1414569A (B9V, V=7.0, K=6.82, d=99pc) were carried
out at the Palomar 200-inch telescope using PALAO the 241-actuators
Adaptive Optics (AO) system installed at the Cassegrain focus (Troy et
al. 2001) and PHARO the near IR camera (Hayward et al. 2001). The
system can reach an averaged Strehl ratio of about 50\% for bright
stars ($m_v<7$) under 1" seeing at near-IR wavelengths, and Strehl
ratio as high as 68\% have been obtained for very good seeing.

HD141569A was observed on May 9, 2001 with a Ks filter ($2.2\,\mu$m)
and a 25\,mas/pixel plate scale.  A coronagraphic
Lyot mask (0.91" in diameter) was used to attenuate the diffraction pattern of the central
star. In addition to the mask, an adequate cold Lyot stop undersizing
the pupil is implemented inside the cryostat of the camera. This
stop is mandatory to remove the unwanted starlight at the edges of the
geometric pupil and to effectively improve the detection of faint
circumstellar material.  The coronagraphic PSF was calibrated in 4 sequences going
back and forth between both the target star and an angularly close
reference star of similar spectral type and magnitude (HD142864, A0V,
V=7.2, K=7.2). The visible magnitude as well as the near-IR magnitude
of this PSF calibrator are optimized to benefit from the same AO
correction but also to obtain a similar signal to noise ratio in the
IR images.  Individual images of both the star and the reference were
separately re-centered and finally coadded to provide an actual
integration time of 1090s although the telescope time required to
perform this thorough calibration amounts to about 2.5 hours. 
Although the target star was observed 10 minutes per sequence (including readout of the detector and sky calibration) it took about 5 to 10 minutes to repoint the telescope in order to accurately re-center the calibrator onto the mask.
The average Strehl ratio was only 16\% during the observation of HD141569 and was measured on the HD141569 companions (7.5" away).

At this point the circumstellar disk is not yet detectable and the
extraction of faint circumstellar material relies on a specific
reduction process.  In order to compare our data with previous HST
observations, we used the same data reduction technique described in
Augereau et al. (1999).  First of all, the data for both the star and
the PSF calibrator were reduced with standard procedure: bad pixels
correction, flat field correction and sky subtraction. Then, before
being subtracted the PSF calibrator need to be recentered and scaled in intensity with respect to the star. We estimate the scaling factor on the ratio
of the target star image to that of the calibrator.  As indicated in
Augereau et al. (1999) the region ranges from 1" to 2" is relatively
free of detected circumstellar dust and the ratio is almost constant
for any orientation angle (Fig. 1).
The global scaling factor is therefore estimated in this area and we
adopt a value of $0.94\pm 0.02$. The uncertainty on the scaling factor
is derived from the pixel to pixel dispersion in the same annulus (see
Fig. 1). The resulting subtracted images are shown on Fig. 2.
\section{Data Analysis}
After subtracting the calibrator, as explained hereabove, it is still difficult to disentangle between circumstellar material since diffraction residuals are dominating the signal around the mask. However, it was possible to unambiguously identify the circumstellar component by comparing our image with the ones obtained using the HST (especially with STIS). Then, to emphasize the disk the diffraction residue was cancelled out on Fig. 2. 
The global elliptical shape of the disk is outlined on Fig. 2b and
several major features in agreement with HST images can be identified.
The ratio between the major and the minor axis is $1.77\pm 0.15$
corresponding to an inclination of $55.6^\circ \pm 3.5^\circ$ from
pole-on assuming a circular disk (in agreement with Mouillet et al. 2001).  
Pieces of two independant rings are
actually discernibles in the disk. The external ring is peaking at
$325$\,AU and was initially detected by Augereau et al. (1999) and
Weinberger et al. (1999). The inner ring is located at $185$\,AU but
is slightly offset by 0.25" ($\approx 2.5\lambda/D$) to the West
as already pointed out by Mouillet et al. (2001). It is however difficult to rely on this value since only 2 pieces of the inner ring are visible on our data and the fitting of an ellipse is somewhat uncertain.
 In addition, an
extended emission detected with STIS is roughly visible to the North
on Fig. 2 but the presence of the 2 companions as well as the
brightness of the background strongly limit the detection in this area. 
The West part of the outer ring (ranging from about $PA=225^\circ$ to
$PA=310^\circ$) is significantly dimmer than the Eastern region by
about a factor of 2.
This can be partly explained by a combination of the inclination of
the disk with respect to the line of sight and anisotropic scattering
properties of the dust grains.
But the present data, like NICMOS2 and STIS images, exhibits some
broken elliptical rings with strong azimuthal asymmetries (especially
in the inner ring).  Mouillet et al. (2001) concluded that a non
axisymmetrical dust distribution was needed to account for the
observed brightness asymmetries.

To obtain more quantitative comparisons with HST data the subtracted
image was azimuthally averaged in several regions to derive the local
surface brightness ($SB$) of the disk. First of all, the image was
deprojected in order to average the pixels at the same
physical distance from the star. 
The surface brightness displayed in mJy/arcsec$^2$ assumes K=6.99 for the central star (as measured in our Ks filter) and a 0\,mag corresponding to a flux density of 667\,Jy.

To evidence the asymmetries of the inner and the outer rings the
surface brightness is plotted on Fig. 3 as a function of the position
angle at respectively $r=1.8"$ and $r=3.2"$ where $r$ denotes the
distance from the star in the deprojected image of the disk. 
Taking $I_{max}$ and $I_{min}$ the maximal and minimal intensity of the azimuthal profiles presented on Fig. 3, the contrast is defined with the following relation:
$$C=\frac{I_{max}-I_{min}}{ I_{max}+I_{min}}$$
The
contrast is as large as 98\% in the inner ring (1.8") between the
bright North-east feature at $PA=20^\circ$ and the depleted South-east
region at $PA=130^\circ$. The outer ring has a lower contrast of 71\%
between the Eastern part ($PA=90^\circ$) and the South-west region
around $PA=230^\circ$.

Figure 4 compares the surface brightness of the disk in the Southern
extension and in the South-Eastern extension where the outer ring is
predominant compared to the inner one. The South-East part appears
dimmer by a factor $1.49\pm 0.35$ in the range $3" \sim 3.5"$. This
disk extension was totally or partially occulted by the wedge on STIS
data and by spider diffraction spikes on NICMOS2 data.
We also found a difference of surface brightness in the southern extension between our
$2.2\,\mu$m data and the HST $1.6\,\mu$m data obtained by Augereau et
al. (1999). The peak of the southern extension on HST image
($SB(1.6\,\mu$m$)$) is $1.33\pm 0.15$ times brighter than in the K
band ($SB(2.2\,\mu$m$)$) as measured on our data (assuming a scaling
factor of 0.94). This difference is discussed in the next section in
terms of scattering properties of the dust grains.

The surface brightness of the inner ring is best evidenced on Fig. 5.
Although the position of the ring is in agreement with the HST data
obtained by Weinberger et al. (1999) it remains difficult to compare
the photometry since the HST image was actually averaged over
$360^\circ$ by Weinberger et al. to derive the surface brightness at
$1.1\,\mu$m. But once our data are processed in the same way we find
that the ring peaks at $0.116\pm0.030$\,mJy/arcsec$^2$ compared to
$0.270\pm0.020\,$mJy/arcsec$^2$ obtained by Weinberger et
al. (1999). The large amount of diffracted light at this angular
separation is mostly responsible for the 25$\%$ uncertainty measured on
the surface brightness at $2.2\,\mu$m.
\section{Implications on disk and grains properties}
\subsection{A possible color-effect}
Since the HD141569 disk is optically thin ($L_{\mathrm{disk}}/L_* =
8.4\times 10^{-3}$, Zuckerman et al. 1995, see also subsection 4.2),
the surface brightness wavelength dependence scales with the star flux
$\Phi^*(\lambda)$ and the scattering cross-section averaged over the
grain size distribution $\langle\sigma_{\mathrm{sca}}(\lambda)\rangle$. We then
tried to investigate whether the observed differences of brightness
between HST data and Palomar images are caused by the star flux only
or reveal a color-effect of the grains. In the following we
investigate the ratio:
\begin{equation}
{\langle\sigma_{\mathrm{sca}}(\lambda_1)\rangle \over
\langle\sigma_{\mathrm{sca}}(\lambda_2)\rangle}={SB(\lambda_1)/\Phi^*(\lambda_1)
\over SB(\lambda_2)/\Phi^*(\lambda_2)}
\end{equation}
with $\lambda_1=1.1\,\mu$m or $1.6\,\mu$m and $\lambda_2=2.2\,\mu$m.
A deviation of this ratio from 1 reveals a non grey scattering
behavior of the grains.

Assuming a B9V spectra for the star we found $\Phi^*(1.6\,\mu$m
$)/\Phi^*(2.2\,\mu$m$)=1.656$. The uncertainty on the ratio
$\langle\sigma_{\mathrm{sca}}(1.6\,\mu$m$)\rangle/\langle
\sigma_{\mathrm{sca}}(2.2\,\mu$m$)\rangle$ at 3.2'' is related to
the surface brightness uncertainty as defined hereabove
($SB(1.6\mu$m$)/SB(2.2\mu$m$)=1.33\pm 0.15$):
\begin{equation}
0.713={1.33-0.15 \over 1.656}<{\langle\sigma_{\mathrm{sca}}(1.6\mu\mathrm{m})
\rangle \over \langle\sigma_{\mathrm{sca}}(2.2\mu\mathrm{m})\rangle}
<{1.33+0.15 \over 1.656}=0.894
\end{equation}
These limits have been compared to numerical simulations in order to
estimate the minimal size of the grains. We assumed first a
collisional differential grain size distribution proportionnal to
$a^{-\kappa}$ with $\kappa=3.5$ and a minimum size $a_{\mathrm{min}}$
(e.g. Hellyer 1970). The use of this size distribution is justified by
the short collision time-scales of the observed dust grains (see next
paragraph). Figure 6 shows the theoretical ratio of the averaged
scattering cross sections $\langle\sigma_{\mathrm{sca}}(1.6\,\mu$m$)
\rangle /
\langle\sigma_{\mathrm{sca}}(2.2\mu\,$m$)\rangle$ as a function of the
minimal grain size $a_{\mathrm{min}}$ for three porosities $P=0$,
$0.5$ and $0.95$ and for typical chemical compositions. Although the
lower limit (0.713) is unhelpful in that case, the upper limit (0.894)
brings some constraints on the grain size distribution. In particular
we found that the minimal size of compact grains ($P=0$) is
$a_{\mathrm{min}}\simeq 0.6\pm 0.2\,\mu$m. This minimal size increases
as the porosity and scales approximatively with $(1-P)^{-1}$:
$a_{\mathrm{min}}\simeq 1.5\pm0.5\,\mu$m for $P=0.5$, and
$a_{\mathrm{min}}\simeq 16\pm9\,\mu$m for $P=0.95$.  However, in this
calculation we used an averaged scaling factor of 0.94 and the grain
size distribution is then no longer constrained if the scaling factor
uncertainty is also considered ($0.94\pm0.2$).
  
The same approach considering now the 1.1\,$\mu$m HST data instead of
the 1.6\,$\mu$m measurements leads to :
\begin{equation}
0.576 <{\langle\sigma_{\mathrm{sca}}(1.1\mu m)\rangle \over
\langle\sigma_{\mathrm{sca}}(2.2\mu m)\rangle}<1.135
\end{equation}
assuming $\Phi^*(1.1\,\mu$m $)/\Phi^*(2.2\,\mu$m$)=2.970$. We then
obtain: $a_{\mathrm{min}}\gtrsim 0.2\times (1-P)^{-1}\,\mu$m,
wich does not improve the accuracy but is in agreement with the
analysis performed at $1.6\,\mu$m and $2.2\,\mu$m. The results are
consistent with grains significantly larger than in the interstellar
medium or in young massive circumstellar disks.  Moreover, these
values do not strongly depend on the assumed size distribution as long
as $\kappa$ is larger than $3$. For small $\kappa$ values, the
scattering cross section is dominated by the large grains that tend to
have a grey scattering behavior not consistent with the observed
color-effect.
\subsection{Basic dynamical considerations}
Assuming circular orbits, we estimate an upper limit on the collision
time-scale of the observed dust grains
by the relation: $t_{\mathrm{coll}}\simeq (2\alpha\tau
\Omega)^{-1}$ where $\Omega =
\sqrt{GM_*/r^3}$ is the Keplerian circular rotation frequency,
$M_* = 2.3\,M_{\odot}$ is the star mass (van den Ancker et al., 1998),
$\tau$ the normal optical thickness of the disk in the near-IR at
distance $r$ and $\alpha$ a constant value depending on grains
optical properties ranging typically between 0.5 and 1 in the regime
of grain sizes considered here. Since the HD141569 disk is not seen
edge-on and is most certainly not geometrically thick (Mouillet et
al. 2001) an estimate of $\tau$ in the near-IR can be derived from the
measured surface brightness $SB(r)$ with the simplified relation:
$\tau(r)\simeq 8\pi r^2 SB(r)/\Phi^*(\lambda)$
(e.g. Appendix A in Augereau et al. 2001). At 2.2\,$\mu$m we derive an
upper limit for the outer ring of $\tau(3.2'') \simeq
3.4^{+0.2}_{-1.1}\times 10^{-2}$ (Southern direction) and $\tau(1.8'')
\simeq 2.3^{+0.1}_{-0.8}\times 10^{-2}$ for the inner ring (N-NE
direction). The positive uncertainty on $\tau(r)$ comes from the
factor of 2 between the East and West part of the disk that can be due
to anisotropic scattering properties of the grains whereas the above
relation assumes grains scatter isotropically. This factor of 2
implies an upper limit on the asymmetry factor $|g|$ for the phase
function of $\sim 0.14$ in a Henyey \& Greenstein (1941) approach
leading to a maximum deviation of $\sim 5\%$ from the isotropic case
for scattering angles close to 90$\degr$. The negative uncertainty
relies also on the optical properties of the grains since $\tau(r)$
can be at most a factor of 1.5 less if the minimum grain size of the
size distribution is small compared to the wavelength
and very porous (at least up to 95$\%$ of porosity). Coming back to
collision time-scales, we obtain $t_{\mathrm{coll}}(3.2'')$ between
$8.7\times 10^3$ and $1.7\times 10^4$\,years and
$t_{\mathrm{coll}}(1.8'')$ between $5.4 \times 10^3$ and $1.1\times
10^4$\,years, a few order of magnitude less than the star age
(8\,Myr)\footnote{Note that $(2\alpha\tau)^{-1}$ represents the
typical number of orbits before a grain on a circular orbit undergoes
a collision and the estimated time-scales correspond to $15\sim 30$
orbits at 3.2'' and $22\sim 44$ orbits at 1.8''}.

The survival of the smallest grains in the system needs then to be
considered. Let us assume first that the disk is free of gas.
In such a case, radiation pressure efficiently blow the smallest
grains out of the system on very short time scales and the HD141569
outer system would then be of ``debris-disk'' type implying the
presence of large bodies replenishing continuously the dust disk.
Therefore, the grains observed in the near-IR and produced by
collisions among large bodies
are gravitationaly linked to the system if their
radiation pressure to gravitational forces ratio is less than 0.5 (see
e.g. Lecavelier 1998 and references therein.) Such dynamical
considerations in a gas free environment give the following
constraints on the minimum grain size in the HD141569 disk:
$a_{\mathrm{min}}\gtrsim 6\,\mu$m if $P=0$,
$a_{\mathrm{min}}\gtrsim 9\,\mu$m if $P=0.5$ and
$a_{\mathrm{min}}\gtrsim 45\,\mu$m if $P=0.95$. According to these
results,
bound grains in the HD141569 system should therefore induce fainter
(or no) color-effect (Fig. 5) than observed. If the measured
color-effect is realistic then grains smaller than the blow-out size
limit in the absence of gas exist in the system. Given the CO J=2-1
brightness temperature measured by Zuckerman et al. (1995) implying a
CO abundance significantly larger than for Main Sequence stars but
faint compared to class II T\,Tauri stars, the assumption on the
absence of gas is in fact probably not correct.
%
\section{conclusion}
HD141569 was used as a fiducial target to assess the capability of the
200-inch telescope in the search of circumstellar disk. The complex
structure of the dusty disk (broken rings and rings offset by 0.25") has been successfully detected at near-IR wavelengths despite a modest Strehl ratio of only 16\%. Therefore, this observation demonstrates the very good performance (comparable to that of the HST) of large ground-based telescopes equipped with
high-order AO systems. Although a few circumstellar disks have been
already imaged from the ground ($\beta$ Pic or HR4796 for instance) in the
near-IR, the 200-inch is to date the largest telescope equipped with a
Lyot coronagraph, thus providing a large angular resolution ($\sim
90mas$ in K band) together with a high dynamic range.  The result of a
search for substellar companion has shown that under good atmospheric
seeing, the detection threshold could be as large as $\Delta m\approx
9mag$ at 0.5", $\Delta m\approx 12mag$ at 1.5" and $\Delta m\approx
14\sim 15mag$ beyond 2".  This capability needs to be intensively
exploited to further discover and characterize other circumstellar
disks.

The many independant data obtain on HD141569 provided us with a
multi-wavelength analysis (visible, near-IR and mid-IR) of this star,
allowing to derive some physical properties of the grains contained in
the disk. In addition to the successfull detection of the disk
reported in this paper we have shown that a careful comparison of
ground-based data and HST data obtained respectively at $1.6\mu$m and
$2.2\mu$m could bring some contraints on the grain size
distribution. In particular, we derive a minimal size of $0.6\pm
0.2\mu m$ for compact grains and a grain size distribution steeper
than $a^{-3}$. Grains are then larger than those found in
circumstellar disks around pre-Main Sequence stars but also smaller
than expected for a Vega-like system.
Therefore, we conclude that gas is probably not fully dissipated in
agreement with the CO detection by Zuckerman et al. (1995).
However, as explained in section 3, the scaling factor between the
target star and the calibrator star is the major source of uncertainty
in that study. Once the scaling factor uncertainty and the porosity
distribution are included in the anlaysis, we end up with a wide range
of grain size.

More accurate results could be probably obtained if the data were
obtained with only one instrument. In that respect, the new generation
of Lyot coronagraphs becoming available on both the Keck telescope and
on the VLT would be certainly helpful for the study of circumstellar
disks.

\acknowledgments{The authors would like to thanks the 200-inch staff
for its efficient support during the observations. This work has been
supported in part by the National Science Foundation Science and
Technology Center for Adaptive Optics, managed by the University of
California at Santa Cruz under cooperative agreement No. AST-9876783,
the french Centre National d'\'Etudes Spatiales (CNES), and is also
contribution xxx from the Lunar and Planetary Institute which is
operated by the Universities Space Research Association under NASA
contract NASW-4574.}


\begin{figure}
\caption{Azimuthally averaged profile of the ratio of the target star image
to that of the calibrator obtained in 4 different regions ($PA=130^\circ$, $234^\circ$, $174^\circ$ and $135^\circ$). The error bars represent the azimuthal intensity dispersion inside each regions. The scaling factor is estimated between 1" and
2" from the central star for which the ratio is almost constant and yields a value of  $0.94\pm 0.02$.}
\end{figure}

\begin{figure}
\epsscale{.75}
\caption{Final image of HD141569 subtracted with a reference star (a/) and
deprojected (b/) assuming an inclination of $55^\circ$. Images c/ and
d/ show the subtraction with a different scaling factor 0.92 and 0.96
respectively $+1\sigma$ and $-1\sigma$ from the averaged value R=0.94
(a/). Two circles are overplotted on the sub-frame b/ to evidence the inner and outer rings. The very bright area at the upper-right corner and the diffraction spikes are provided by the T Tauri companions.}
\end{figure}

\begin{figure}
\caption{Variation of the surface brightness across the inner ring (top) and
the outer ring (bottom) as a function of the Position Angle. The
contrast of the asymetries is as large as 98\% in the inner ring and
71\% in the outer ring.}
\end{figure}

\begin{figure}
\caption{Surface brightness profile of the Southern region ($PA=180^\circ\pm
10^\circ$) and the South-East region ($PA=130^\circ\pm 10^\circ$)
compared to the surface brightness of the disk obtained with NICMOS at
$1.1\mu m$ (Weinberger et al. 1999) and at $1.6\mu m$ (Augereau et
al. 1999). The captioned image shows the azimuthally averaged regions
around the star. The brightness profile at $1.1\mu m$ is a rough
estimation of the Fig. 2 displayed in Weinberger et al. (1999). The
profile at $1.6\mu m$ is obtained with real data obtained by Augereau
et al. (1999).}
\end{figure}

\begin{figure}
\caption{Surface brightness profile of the North-East region ($PA=18^\circ\pm
10^\circ$) and the South-West region ($PA=215^\circ\pm 10^\circ$)
compared to the surface brightness of the disk obtained with NICMOS at
$1.1\mu m$ (Weinberger et al. 1999) and at $1.6\mu m$ (Augereau et
al. 1999). The captioned image shows the azimuthally averaged regions
around the star. The brightness profile at $1.1\mu m$ is a rough
estimation of the Fig. 2 displayed in Weinberger et al. (1999). The
profile at $1.6\mu m$ is obtained with real data obtained by Augereau
et al. (1999).}
\end{figure}

\begin{figure}
\caption{Numerical simulations of the theoretical ratios
$\langle\sigma_{\mathrm{sca}}(1.1\,\mu$m$)\rangle/
\langle\sigma_{\mathrm{sca}}(2.2\,\mu$m$)\rangle$
and $\langle\sigma_{\mathrm{sca}}(1.6\,\mu$m$)\rangle/
\langle\sigma_{\mathrm{sca}}(2.2\,\mu$m$)\rangle$ as a function of the
minimal grain size $a_{\mathrm{min}}$, for three porosities $P=0$,
$P=0.5$ and $P=0.95$ and assuming a grain size distribution following
a -3.5 power law. The following chemical compositions have been
assumed~: graphite, amorphous and crystalline silicates for $P=0, 0.5,
0.95$, mixed with organic refractories and/or water ice (10$\%$ of the
vaccum due to porosity) for non compact grains ($P=0.5, 0.95$).  The
upper and lower limits derived from Eq. 2 and Eq. 3 are also
overplotted.}
\end{figure}
\end{document}